\documentstyle[prb,aps,graphicx,multicol]{revtex}

\begin{document}
\draft
\title{Resonances in a two-dimensional electron waveguide with a
single delta-function scatterer}
\author{Daniel Boese\cite{new1}, Markus Lischka\cite{new2} and
L.E.~Reichl}
\address{Center for Studies in Statistical Mechanics and Complex
Systems, The University of Texas at Austin, Austin, Texas 78712}
\date{\today}
\maketitle
\begin{abstract}
We study the conductance properties of a straight two-dimensional
electron waveguide with an $s$-like scatterer modeled by a single
$\delta$-function potential with a finite number of modes.  Even such
a simple system exhibits interesting resonance phenomena.  These
resonances are explained in terms of quasi-bound states both by using
a direct solution of the Schr\"odinger equation and by studying the
Green's function of the system.  Using the Green's function we
calculate the survival probability as well as the power absorption and
show the influence of the quasi-bound states on these two quantities.
\end{abstract}

\begin{multicols}{2}

\narrowtext

\section{Introduction}
\label{sec:intro}
Two-dimensional electron systems have been studied extensively over
the past few years, both because it became feasible to construct such
systems for example at a GaAs/GaAlAs interface at low temperatures and
because the conductance was shown to be directly related to the
transmission properties of the system.  This relation is known as
Landauer's formula, \cite{fisher:81,stone:88,baranger:89}
\begin{equation}
    \Gamma = \frac{e^{2}}{h} T,
    \label{eq:landauer}
\end{equation}
where $\Gamma$ denotes the conductance and $T$ the full transmission
function (spin degrees of freedom are neglected). 
Eq.~(\ref{eq:landauer}) results in a quantized conductance for
straight channels.  If one adds a single attractive $\delta$-function
scatterer to such a straight waveguide, the combined effect of the
scatterer itself and the backscattering off the walls leads to
interesting phenomena.  This model was first suggested by Datta et
al.\cite{datta:87} and later discussed by Bagwell et
al.\cite{bagwell:90,bagwell:92} In the following we will take a closer
look at the resonance phenomena this system produces.

\section{Direct solution of the Schr\"odinger equation}
\label{sec:direct}
\begin{figure}
    \centerline{\includegraphics[scale=0.5]{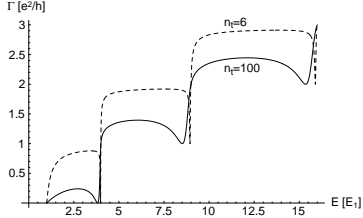}}
    \caption{Total conductance for a waveguide with a $\delta$-function
    scatterer of strength $\gamma = - 7 \, \mathrm{feV} \,
    \mathrm{cm}^{2}$.  The conductance goes to zero just before the second
    subband opens.}
    \protect\label{fig:conductance}
\end{figure}

\begin{figure}
    \centerline{\includegraphics[scale=0.33]{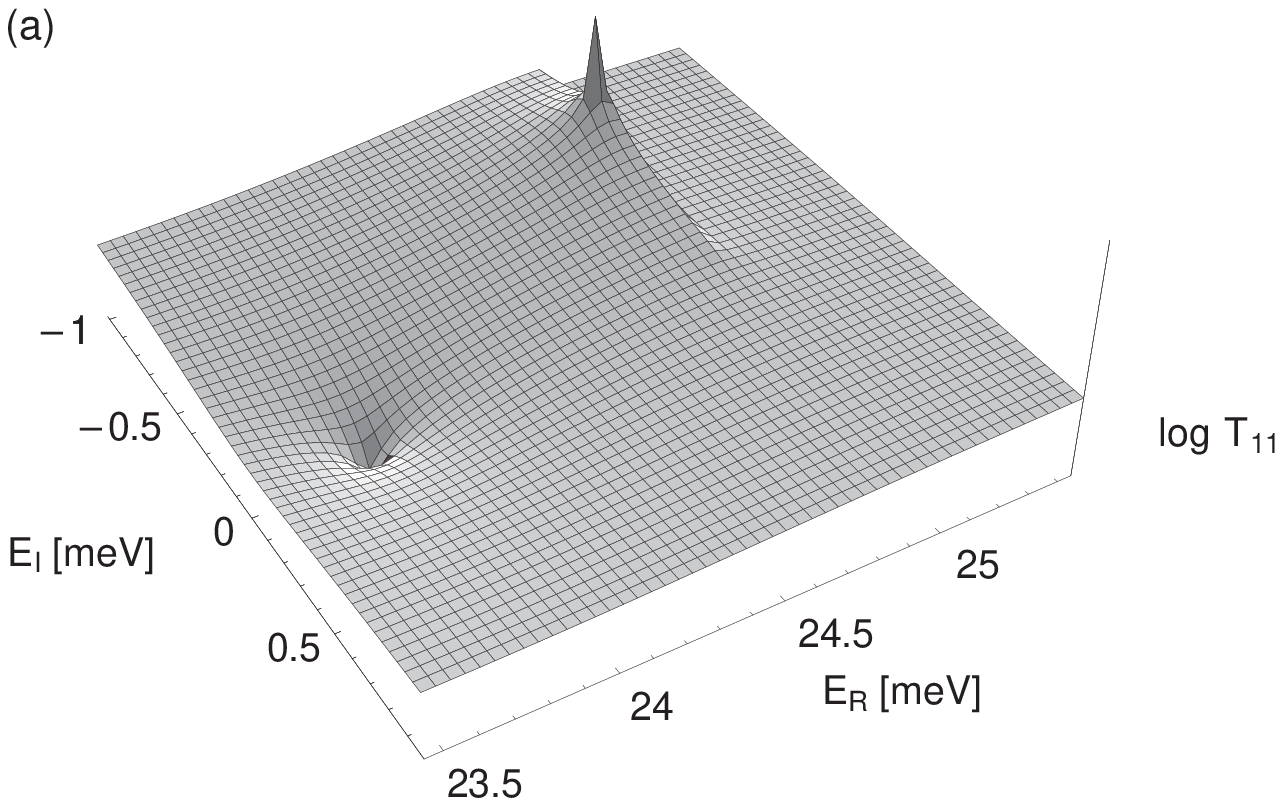}
    \includegraphics[scale=0.33]{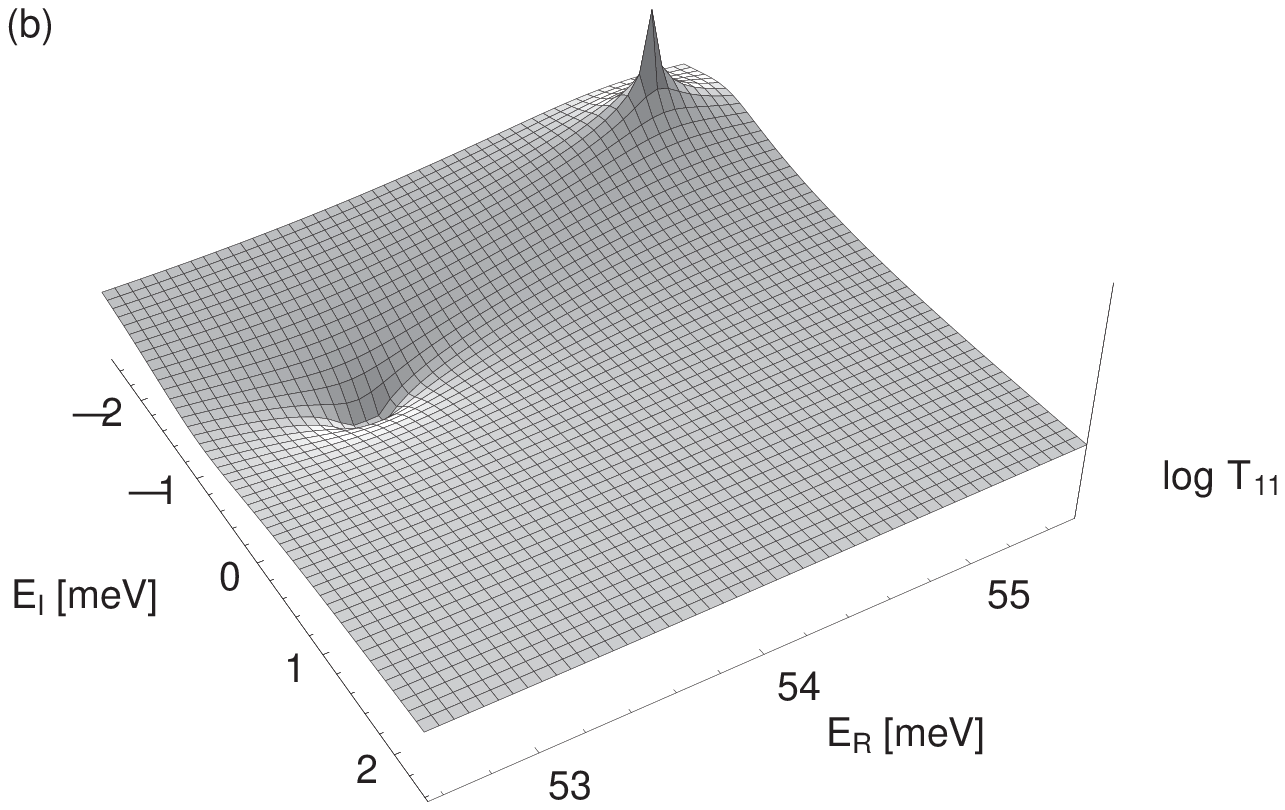}}
    \caption{Transmission poles for $\gamma = - 7 \, \mathrm{feV} \,
    \mathrm{cm}^{2}$.  The poles just below the second (a) and third (b)
    subband edge are shown together with the corresponding zeros which
    appear as dips in these logarithmic plots.  The zero in (b) is off the
    real axis as the transmission decreases to 1 instead of 0 at this
    resonance (cf.\ Fig.~\ref{fig:conductance}).}
    \protect\label{fig:transpole}
\end{figure}

Let us first obtain a solution of the Schr\"odinger equation for an
electron in a two-dimensional waveguide with a $\delta$-function
scatterer.  The Hamiltonian is given by
\begin{equation}
    H = \frac{p^{2}}{2 m} + V(x,y) + V_{c}(y).
    \label{eq:H}
\end{equation}
$V_{c}$ represents a confinement potential restricting the movement
of the electron to the range $0 < y < D$.\cite{hopot} The attractive
scattering potential is given by
\begin{equation}
    V(x,y) = \gamma \, \delta (x) \delta (y-y_{0}) \qquad (\gamma < 0).
    \label{eq:Vdelta}
\end{equation}
We can now expand any stationary solution $\psi_{E}(x,y)$ of the
Schr\"odinger equation, $H \psi_{E}(x,y) = E \psi_{E}(x,y)$, in a
Fourier series with $x$-dependent expansion coefficients using the
complete set of transversal modes,
\begin{equation}
    \psi_{E} \left( x,y \right) = \sum_{n=1}^{\infty} c_{n} (x) \chi_{n}
    (y),
\end{equation}
where $\chi_{n}(y) = \sqrt \frac{2}{D} \sin \frac{n \pi}{D} y$. 
Inserting this series into the Schr\"odinger equation and employing
orthogonality of the transversal modes we obtain a set of coupled
equations,
\begin{equation}
    \frac{\partial^{2}}{\partial x^{2}} c_{m} \left( x \right) + k_{m}^{2}
    c_{m} (x) = \sum_{n} M_{mn} c_{n}(x) \delta(x),
    \label{eq:cmpde}
\end{equation}
where $M_{mn} = \frac{4 m \gamma}{D \hbar^{2}} \sin \frac{m \pi
y_{0}}{D} \sin \frac{n \pi y_{0}}{D}$ denote the coupling constants
and $k_{n} = \sqrt{\frac{2 m}{\hbar^{2}} E - \frac{n^{2}
\pi^{2}}{D^{2}}}$ the wavevector (${\mathrm{Im}} \, k_{n} > 0$).  Away
from the scatterer, $x \neq 0$, the wavefunction must have the
free-electron form,
\begin{equation}
    c_{n}(x) = \left\{ \begin{array}{ll}
        A_{n} e^{i k_{n} x} + B_{n} e^{-i k_{n} x} &\quad x < 0 \\
        C_{n} e^{i k_{n} x} + D_{n} e^{-i k_{n} x} &\quad x > 0
    \end{array} \right.
    \label{eq:cnsol}
\end{equation}
As $\psi$ must be continuous at $x = 0$ and its derivative must have a
finite jump there, the same conditions must hold for the expansion
coefficients $c_{n}(x)$.  Thus using these two conditions on
Eq.~(\ref{eq:cmpde}) with the ansatz in Eq.~(\ref{eq:cnsol}) yields
\begin{eqnarray}
    A_{n}+B_{n} &&= C_{n}+D_{n},
    \label{eq:set1} \\
    i k_{n} \left( C_{n} - D_{n} \right) - && i k_{n} \left( A_{n} - B_{n}
    \right)
    \nonumber \\
    &&= \sum_{m} M_{nm} \left( A_{m} + B_{m} \right).
    \label{eq:set2}
\end{eqnarray}
If $\psi$ is an evanescent mode, we can set $k_{n} = i \kappa_{n}$ and
must require $A_{n} = 0$ and $D_{n} = 0$ to have a normalizable
wavefunction.  The transmission coefficient for propagating modes is
then defined as $T_{mn} = \frac{k_{n}}{k_{m}} \frac{\left| C_{n}
\right|^{2}}{\left| A_{m} \right|^{2}}$ and the total transmission
function as
\begin{equation}
    T(E) = \sum_{mn \atop \mathrm{(prop.)}} T_{mn}
    \label{eq:totaltrans}
\end{equation}
where the sum extends over all propagating modes.  The conductance is
finally calculated using Eq.~(\ref{eq:landauer}).  The set of
Eqs.~(\ref{eq:set1}) and (\ref{eq:set2}) can only be solved
numerically for a finite number of modes.  We use the parameters $D =
300 \, \mathrm{\AA}$ for the width of the channel and $y_{0} =
\frac{5}{12} D$ for the transversal position of the scatterer, the
mass $m = 0.067 m_{e}$ as the effective mass of an electron in GaAs
and a scatterer strength of $\gamma = - 7 \, \mathrm{feV} \,
\mathrm{cm}^{2}$.  With these parameters the first four energy
subbands (transversal modes) open up at $E_{1}=6.24 \, \mathrm{meV}$,
$E_{2}=24.94 \, \mathrm{meV}$, $E_{3}=56.12 \, \mathrm{meV}$ and
$E_{4}=99.78 \, \mathrm{meV}$.  Numerical results of
Eq.~(\ref{eq:landauer}) for a total number of modes $n_{t} = 6$ and
$n_{t} = 100$ are shown in Fig.~\ref{fig:conductance}.  It shows the
expected step-like behavior at every energy subband edge. 
Furthermore, just before a new subband opens up, it shows an
interesting dip.  The waveguide blocks transmission in the lower mode
just before a new higher mode opens up.  It becomes completely opaque
just before the second mode opens up.  The drop in the conductance to
the $n$-th level just before the next higher mode $n+1$ opens up is in
fact a cumulative effect of all transmitting modes $1$ through $n$ as
can be seen from the individual transmission
coefficients.\cite{bagwell:90} These dips correspond to a resonance
structure of the system.  They can be attributed to quasi-bound states
of the system with a finite lifetime, represented by a wavefunction of
the form\cite{landau3}
\begin{equation}
    \psi ({\mathbf{x}},t) = \psi ({\mathbf{x}}) e^{E_{R} t / i \hbar}
    e^{E_{I} t / \hbar},
    \label{eq:qbstate}
\end{equation}
where $E_{R}$ and $E_{I}$ denote the real and imaginary part
respectively.  These states are characterized by having a scattered
wave even without an incident wave, i.e.\ $1/T_{mn} = 0$.  We thus
look for corresponding poles of the transmission function near each of
these resonances.  Figures~\ref{fig:transpole} (a) and (b) show the
poles corresponding to the first two dips of
Fig.~\ref{fig:conductance} confirming our assertion.  The poles are
always located in the analytically continued second sheet of the
square root function (${\mathrm{Im}} \, k_{n} < 0$).  For the plots in
Fig.~\ref{fig:transpole} the square root function was chosen to have
its branch cut on the negative imaginary axis which is visible as a
discontinuity in the plots (It is thus the square root function with
values on the first, ``physical'' sheet in the first through third
quadrant of the complex energy plane and the square root function with
values on the second sheet in the fourth quadrant that is used in
Fig.~\ref{fig:transpole}).

\section{Green's function approach}
\label{sec:green}
\begin{figure}
    \centerline{\includegraphics[scale=0.33,angle=270]{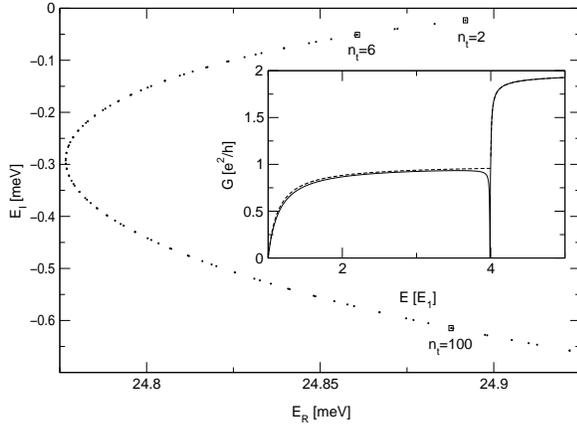}}
    \caption{Poles of G in the complex energy plane depending on the
    number of total modes $n_{t}$.  The pole for $n_{t}=100$ is located at
    $E = 24.888 - 0.615 i \, \mathrm{meV}$ and corresponds to the one
    shown in Fig.~\ref{fig:transpole} (a).  Inset: Conductance for $n_t=2$
    (solid) and $n_t=$ number of propagating modes (dashed).  For $E>E_2=4
    E_1$ both are identical.}
    \protect\label{fig:Gpoles}
\end{figure}

The resonances observed in Section~\ref{sec:direct} can be determined
from the Green's function of the system as well.\cite{taylor} We thus
try to find the exact Green's function of the Hamiltonian $H$.  This
on the one hand allows for a direct and exact calculation of the
quasi-bound state energies as poles of the Green's function (on the
second sheet).  On the other hand, the Green's function calculation is
less compute-intensive and thus allows the calculation of both the
survival probability and the power absorption of the waveguide (see
Sec.~\ref{sec:survival} and \ref{sec:power}).  We first obtain the
solution of the Green's function equation for the free waveguide with
$H_{0} = \frac{p^{2}}{2 m} + V_{c}(y)$, \cite{morse1}
\begin{eqnarray}
    \lefteqn{G^{0, R/A} \left( x,y,x',y',E \right) = }\qquad
    \label{eq:G0sol} \\
    && \pm \sum_{n = 1}^{\infty} \frac{2}{D} \sin \left( \frac{n \pi}{D} y
    \right) \sin \left( \frac{n \pi}{D} y' \right) \frac{2 m}{\hbar^{2}}
    \frac{e^{\pm i k_{n} \left| x-x' \right|}}{2 i k_{n}}.
    \nonumber
\end{eqnarray}
$k_{n}$ is given as in Section~\ref{sec:direct}.  The retarded
solution ($+$) is denoted by $G^{0, R}$, the advanced solution ($-$)
by $G^{0, A}$.  For the special potential $V(x,y)$ the integral
equation for the full Green's function can be solved to yield
\cite{ludviksson:87,bagwell:90b}
\begin{eqnarray}
    \lefteqn{G^{R/A} \left( x,y,x',y',E \right) = G^{0, R/A} \left(
    x,y,x',y',E \right)}\qquad\qquad
    \label{eq:Gsol} \\
    && + \frac{G^{0, R/A} \left( x,y,0,y_{0},E \right) G^{0, R/A} \left(
    0,y_{0},x',y',E \right) } {1/\gamma - G^{0, R/A} \left(
    0,y_{0},0,y_{0},E \right) }.
    \nonumber
\end{eqnarray}
The transmission is solely determined by the retarded Green's
function.\cite{fisher:81} We can thus read off the condition for the
poles to be
\begin{equation}
    \frac{1}{\gamma} - \sum_{n} \frac{4 m}{D \hbar^{2}} \left( \sin
    \frac{n \pi}{D} y_{0} \right)^{2} \frac{1}{2 i k_{n}} = 0.
    \label{eq:Gpolecond}
\end{equation}
It can be explicitly shown from Eq.~(\ref{eq:Gpolecond}) that the
poles are always located on the second sheet and hence the branch cut
has to be chosen as described in Section~\ref{sec:direct}.  Evaluating
Eq.~(\ref{eq:Gpolecond}) again numerically for $n_{t}=6$ and
$n_{t}=100$ modes gives the pole locations in Table~\ref{tab:Gpoles}. 
Apart from the quasi-bound states the attractive $\delta$-function
scatterer also exhibits one single bound state.  The pole locations
agree exactly with the results from Section~\ref{sec:direct} thus
confirming the interpretation of the resonance phenomena to
quasi-bound states.  For the discussed $\delta$-function impurity, the
poles depend sensitively on the number of modes included in the
computation.  The typical behavior of the pole location vs.  number of
modes is shown in Fig.~\ref{fig:Gpoles} for the pole just below the
second subband edge.  The plot demonstrates that the location of the
poles and thus the conductance properties of the discussed system are
not converging with an increasing number of modes.  It can also be
seen that they are systematically shifted to greater imaginary values,
displayed in the broadening of the resonance.

The sum in Eq.~(\ref{eq:Gpolecond}) is in fact logarithmically
diverging as each term in the sum is positive and approximately
proportional to $1/n$ for large $n$.  The obtained results thus have
to be always discussed for a specific number of modes.  Nevertheless,
the numerical results are in qualitative agreement with the behavior
of an $s$-like scatterer as discussed by Kunze and
Lenk.\cite{kunze:92} They thus still serve their purpose as a useful
model if the $\delta$-function scatterer with a finite number of modes
is interpreted as an $s$-like scatterer with finite width $D/n_{t}$ in
$y$-direction instead of a true $\delta$-function scatterer.

Before proceeding let us review the resonances in the context of the
last two sections.  They are formed by all modes of the system
collectively.  Both propagating and evanescent modes (and therefore
inter-channel coupling) are necessary.  The evanescent modes are
needed to build up a bound or quasi-bound state, whereas the
propagating modes probe that state and therefore display the structure
of the resonance.  This can be seen clearly in Fig.~\ref{fig:Gpoles}
which shows the position of the pole causing the resonance just below
the threshold of the second propagating mode.  Keeping only two modes,
$n=1$ which is propagating and $n=2$ which is extended and evanescent
just below the threshold of the $n=2$ propagating mode, is necessary
to create the pole and hence the resonance.  This can be seen clearly
in the inset of Fig.÷\ref{fig:Gpoles} where we show the conductance
with only the $n=1$ propagating mode and the conductance when both the
$n=1$ and $n=2$ modes are present.  There is no resonance when only
the $n=1$ mode is present.  Thus, excluding the evanescent modes leads
to unphysical kinks in the conductance near the threshold.  The main
contribution to any given resonance stems from the evanescent mode
that is about to become propagating, because it has an infinite decay
range at the band edge.  The dip as well as the position of the pole
are only slightly modified if more than this evanescent mode are
included, but neither the dip nor the pole exist without evanescent
modes.  The remaining evanescent modes only alter the width and
position of the resonances and hence play a minor role.  It is
interesting to note that when more than one scatterer is present,
evanescent modes play an even more important role because they can
localize coherently over more than one scatterer.  The case of many
scatterers has been discussed for waveguides in
Ref.~\onlinecite{kumar:90,kumar:91,kumar:91b,bandyopadhyay:91} and for
closed quantum systems in Ref.~\onlinecite{wendler:95}.

\section{Survival Probability}
\label{sec:survival}
\begin{figure}
    \centerline{\includegraphics[scale=0.5]{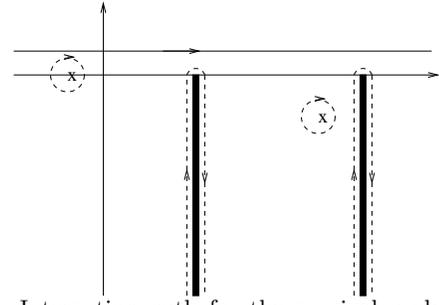}}
    \caption{Integration path for the survival probability before (solid)
    and after deforming the contour (dotted) in the complex $z$-plane.}
    \protect\label{fig:intpath}
\end{figure}

\begin{figure}
    \centerline{\includegraphics[scale=0.33]{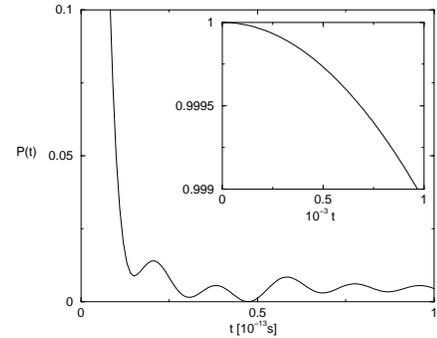}}
    \caption{The survival probability $P(t)$ as a function of time.  The
    inset illustrates the Zeno effect for short times.}
    \protect\label{fig:survival}
\end{figure}

Up to now the Green's function only allowed for a more compact
formulation.  We now demonstrate that it will also give new insight on
other quantities like the survival probability.  The survival
probability is defined as
\begin{equation}
    P(t)=\left| \left\langle \psi (t) \right| \left.  \!\psi_i
    \right\rangle \right|^2,
    \label{eq:survdef}
\end{equation}
where $\left| \psi_i \right\rangle$ is the initial state at time $t=0$
and $\left| \psi (t) \right\rangle$ is the propagated state, which can
be computed from
\begin{eqnarray}
    \psi({\mathbf{x}},t)&=&\lim_{\epsilon \rightarrow 0} \left[
    \frac{i}{2\pi} \int_{-\infty}^\infty \!  d{\mathbf{x'}}
    \int_{-\infty+i\epsilon}^{\infty+i \epsilon} dz \right.
    \nonumber \\
    && \left.  e^{-\frac{izt}{\hbar}}G^R({\mathbf{x}},{\mathbf{x'}},z) \,
    \psi_i({\mathbf{x'}}) \right], \quad {\mathbf{x}} = (x,y),
    \label{eq:propstate}
\end{eqnarray}
where the Green's function enters explicitly into the calculation. 
The integration runs in the complex plane as shown in
Fig.~\ref{fig:intpath} (solid line).  We can deform the contour and
perform the integration as indicated by the dotted lines.  In this
case we pick up poles as well as integrations along our choice of
branch cuts.  For the ballistic case however, the Green's function
does not have any poles and only the integrations along the cuts,
performed on different sheets, remain.  We now demonstrate this case
explicitly, for which we write the initial state in the following form
$\psi_i({\mathbf{x}})=X(x) Y(y)$.  Both $X(x)$ and $Y(y)$ are
localized and do not contain an explicit $z$-dependence.  For the
integration along the $n$-th cut we define
\begin{eqnarray}
    K_n^{0,R}({\mathbf{x}},{\mathbf{x'}},t) =&& \\
    \sum_{n'} \frac{m}{\hbar^2} && \chi_{n'}^\ast (y) \chi_{n'} (y')
    \int_{C_n} \!\!\!dz \, e^{-\frac{izt}{\hbar}}\frac{e^{i k_{n'}
    |x-x'|}}{ik_{n'}}.
    \nonumber
\end{eqnarray}
Although the integration runs on two different sheets, it is only
relevant for the $n$-th square root, i.e.\ for $k_n$.  Hence only one
term contributes in the sum and we obtain
\begin{eqnarray}
    K_n^{0,R}({\mathbf{x}},{\mathbf{x'}},t) =&& \\
    \frac{1}{i e^{i\frac{3}{4} \pi}} && \sqrt{\frac{2m \pi}{\hbar
    t}}e^{\frac{im(x-x')^2}{2 \hbar t}}e^{\frac{i E_n t}{\hbar}}
    \chi_n^\ast (y) \chi_n (y').
    \nonumber
\end{eqnarray}
Putting everything together we find for the survival probability of
the ballistic wire
\begin{eqnarray}
    P(t)&=& \left| \frac{1}{i e^{i\frac{3}{4} \pi}} \sqrt{\frac{2m
    \pi}{\hbar t}} \sum_n e^{\frac{i E_n t}{\hbar}} \, \left| \int \!\! 
    dy Y(y) \chi_n(y) \right|^2\right.
    \nonumber \\ 
    & &\left.  \int \!\!dx \!\int \!\!dx' \, e^{\frac{im(x-x')^2}{2 \hbar
    t}} X^\ast (x) X(x') \right|^2 .
\end{eqnarray}
In Fig.~\ref{fig:survival} we plot the survival probability for an
initial state which is Gaussian-like localized in the $x$-direction
and has a mode expansion for the transverse part, i.e.\ it couples
equally strong to all channels.  We distinguish between short, medium
and long time behavior.  For the short time behavior we observe the
Zeno effect,\cite{Zeno} i.e.\ an initially non-decaying behavior.  In
the medium time regime we observe oscillations of all contributing
channels in the decaying probability.  The long time behavior is
dominated by an $1/t$ behavior.

For the case of a $\delta$-function impurity in the wire we pick up
poles which contain terms like $e^{i E_{\mathrm{QBS}} t/\hbar}$ and
therefore justify Eq.~(\ref{eq:qbstate}).  Experimentally they can be
detected by means of a Fourier transform and thus provide a tool to
probe the complex spectrum.  In addition we get contributions from the
cuts, which do not cancel out, but are not dominating either.

\section{Power Absorption}
\label{sec:power}
Another quantity of interest is the power absorption $\left\langle P
\right\rangle$ of a quantum wire, which can be calculated from
microscopic theory.  It is related to the AC conductance $G(\omega)$
via $\left\langle P \right\rangle = E_{\mathrm{rms}}^2 L^2 G(\omega)$
and hence allows us to make a straight connection between quasi-bound
states and dips in the conductance.

\begin{figure}
    \centerline{\includegraphics[scale=0.33]{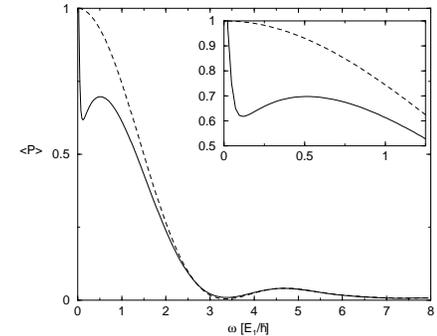}}
    \caption{Power absorption as a function of $\omega$ in units of
    $\frac{2h}{e^2E_0^2L^2}$ for the ballistic (dashed) and the impurity
    wire (solid).  The Fermi energy is close to the quasi-bound state
    energy.}
    \protect\label{fig:power}
\end{figure}

An electric AC field of amplitude $E_{0}$ is applied to a region of
length $L$, which is symmetric around the scatterer.  It can be
shown\cite{powerkramer} that in linear response the power absorption
of the ballistic wire behaves for small $\omega$'s like
\begin{equation}
    \left\langle P \right\rangle = \frac{L^2 e^2
    E_0^2}{2h}\sum_{n=1}^{n_C}\left(\frac{\sin\!\left(\frac{m\omega
    L}{2k'_+ \hbar}\right)}{\left(\frac{m\omega L}{2k'_+
    \hbar}\right)}\right)^2
    \label{eq:ballpower}
\end{equation}
with $k'_{+}=\sqrt{k_{F}^{2}-\left( \frac{n \pi}{D} \right)^{2}}$. 
The quasi-bound states have a profound influence on the power
spectrum.  Starting from the microscopic expression $P(t)=\int \! 
d{\mathbf{r\, E(r}},t) \cdot \left\langle
{\mathbf{j(r}},t)\right\rangle$ one can show\cite{powerna} that the
quasi-bound states give Lorentzian-like contributions to the power
absorption in these systems.  However it is not clear whether these
are positive or negative.  We calculate the power absorption from
\begin{eqnarray}
    \left\langle P \right\rangle &=& \frac{-\hbar \pi}{2} \int
    \!\!d{\mathbf{r}}\int \!\!d{\mathbf{r'}} \,E(x)E(x')
    \int_{-\infty}^{\infty}\!\!d\tilde{E}_1
    \nonumber \\ 
    & & \times \frac{f(\tilde{E}_1)-f(\tilde{E}_1+\hbar \omega)}{-\hbar
    \omega} \, \left\langle\tilde{E}_1\right| J_x({\mathbf{r}})
    \left|\tilde{E}_1 +\hbar \omega \right\rangle
    \nonumber \\ 
    & & \times \left\langle\tilde{E}_1+ \hbar \omega \right|
    J_x({\mathbf{r'}}) \left| \tilde{E}_1 \right\rangle .
    \label{eq:powereval}
\end{eqnarray}
The microscopic current elements are evaluated in a scattering state
basis.  Our numerical results are shown in Fig.~\ref{fig:power}.  The
applied field is of length $L=100 \, \mathrm{nm}$, the scattering
states are normalized on $L'=4000 \, \mathrm{nm}$.  We clearly
reproduce the analytical result for small $\omega$, which holds
surprisingly well even for larger values.  For the case of a
$\delta$-function impurity the spectrum is still dominated by the
ballistic background, which is not surprising because the impurity is
strongly localized.  The influence of the quasi-bound states can also
be seen, however it is quite smaller than for the system discussed by
Na and Reichl.\cite{powerna} In contrast to their system we see dips
rather than peaks.  This could be due to the fact that we excite into
a quasi-bound state from a continuum state, which is why the dip
shifts as a function of the Fermi energy.  Na and {Reichl} however do
excite from one quasi-bound state into another so that the signs
cancel each other.  In our system these resonances are too small to be
observed, which can be attributed to the fact that the perturbation
causing the quasi-bound state is much smaller (a $\delta$-function
with a finite number of modes) than the large cavity in their
system.\cite{locstate} On the other hand we can now make a clear and
unambiguous connection between the dips in the DC conductance and the
quasi-bound states.

\section{Conclusion}
\label{sec:conclusion}
We have considered a simple model for an electron waveguide.  The
conductance was calculated via Landauer's formula and shown to exhibit
resonance phenomena.  These resonances were attributed to quasi-bound
states of the system by both looking at poles of the transmission
coefficients and the Green's function.  We have shown that the
delta-function potential together with a finite number of modes models
an $s$-like scatterer.  Furthermore, we have demonstrated how these
quasi-bound states influence the survival probability and the power
absorption of the system.

\section*{Acknowledgments}
The authors wish to thank the Welch Foundation, Grant No.~1052 and DOE
Contract No.~DE-FG03-94ER14405 for partial support of this work.  Two
of the authors (D.B., M.L.) gratefully acknowledge financial support
from the German National Merits Scholarship Foundation during their
stay in Austin.

%

%
\begin{table}[tbp]
    \centering
    \caption{Poles of $G$ for a scatterer strength of $\gamma = -7 \,
    \mathrm{feV} \, \mathrm{cm}^{2}$ and a total number of modes $n_{t} =
    6$ and $n_{t} = 100$}
    \label{tab:Gpoles}
    \begin{tabular}{lll}
        Pole & $E$ [meV] $(n_{t} = 6)$ & $E$ [meV] $(n_{t} = 100)$ \\ \hline
        0 & $4.676$ & $-8.656$ \\
        1 & $24.861 - 0.051 i$ & $24.888 - 0.615 i$ \\
        2 & $55.804 - 0.144 i$ & $55.117 - 1.961 i$ \\
        3 & $99.205 - 0.262 i$ & $98.073 - 2.561 i$ \\
    \end{tabular}
\end{table}

\end{multicols}


\begin{references}
\bibitem{hopot}
The well potential has been chosen for numerical simplicity.  In
principle, a model closer to the experiment, like the harmonic
oscillator potential, can be chosen as well.
\bibitem{locstate}
In fact, the wavefunction of the quasi-bound state of the
$\delta$-function perturbation is highly localized around the
scatterer.
\end{references}
\end{document}